\documentclass[ reprint,
groupedaddress,
 amsmath,amssymb,
 aps,
prl,
]{revtex4-2}

\usepackage{appendix}
\usepackage{graphicx}
\usepackage{dcolumn}
\usepackage{bm}
\usepackage{hyperref}
\usepackage[mathlines]{lineno}
\usepackage{mathtools}
\usepackage{xcolor}
\usepackage{titlesec}

\newcommand{\rone}{r_{\rm s}}
\newcommand{\rtwo}{r_{\rm e}}
\newcommand{\rthree}{r_{\rm g}}
\newcommand{\reff}{\eta}

\titleformat{\section}[runin]{\itshape}{}{0em}{}[---]
\titleformat{\subsection}[runin]{\itshape}{}{0em}{~~}[---]

\begin{document}

\title{Gravitational Gertsenshtein–Zel'dovich mechanism for the Association between GW190425 and FRB 20190425A}

\author{Shao-Qin Wu$^{1,7,8}$}
\thanks{co-first authors}
\author{Jing-Rui Zhang$^{1,7,8}$}
\thanks{co-first authors}%

\author{Rong-Gen Cai$^{3}$}
\email{caironggen@nbu.edu.cn}
\author{Bing Zhang$^{4,5,6}$}
\email{bzhang1@hku.hk}
\author{Yun-Long Zhang$^{2,1}$}
\email{zhangyunlong@nao.cas.cn}

\affiliation{$^{1}$School of Fundamental Physics and Mathematical Sciences,  Hangzhou   Institute for Advanced Study, University of Chinese Academy of Sciences,, Hangzhou 310024, China.}

\affiliation{$^{2}$National Astronomical Observatories, Chinese Academy of Sciences, Beijing, 100101, China}

\affiliation{$^{3}$
Institute of Fundamental Physics and Quantum Technology, Ningbo University, Ningbo 315211, China}

\affiliation{$^{4}$
The Hong Kong Institute for Astronomy and Astrophysics, University of Hong Kong, Pokfulam Roads, Hong Kong, China
}

\affiliation{$^{5}$
Department of Physics, University of Hong Kong, Pokfulam Road, Hong Kong, China
}

\affiliation{$^{6}$
Department of Physics and Astronomy, University of Nevada Las Vegas, Las Vegas, NV 89154, USA
}

\affiliation{$^{7}$ Institute of Theoretical Physics, Chinese Academy of Sciences, Beijing 100190, China}

\affiliation{$^{8}$Taiji Laboratory for Gravitational Wave Universe (Beijing/Hangzhou), University of Chinese Academy of Sciences, Beijing 100049, China}

\date{\today}

\begin{abstract}
The temporal and spatial coincidence between the gravitational wave (GW) event GW190425 and the fast radio burst (FRB) event FRB 20190425A raises the intriguing possibility of a physical connection between the two. The widely discussed possibility invoking the collapse of a supermassive neutron star as the merger product suffers the inconsistency between the model prediction and the measured inclination angle of the system. Here, we propose a novel physical mechanism to account for the association. We envisage a magnetar located at about 2.5 light hours away from the binary neutron star merger site. The kiloherz GWs generated by the merger are converted into kiloherz electromagnetic (EM) radiation via the Gertsenshtein–Zel'dovich (GZ) effect near the magnetar. Subsequent inverse Compton scattering off the kilohertz EM waves by relativistic particles generates the observed gigahertz FRB emission. Our calculation reveals that, with appropriate parameter choices, the properties of FRB 20190425A can be reproduced.

\end{abstract}

\maketitle
\allowdisplaybreaks

\section{Introduction}
Fast radio bursts (FRBs) \citep{Lorimer:2007qn, Thornton:2013iua} are bright and millisecond-duration radio bursts that originate from cosmological distances, whose origin is still mysterious \citep{Cordes:2019cmq, Platts:2018hiy, Zhang:2022uzl}. Observations suggest that at least some and probably most FRBs are repeating events \citep{Spitler:2016dmz,2021ApJS..257...59C}, suggesting a non-catastrophic engine such as magnetars. Nonetheless, the possibility that a small fraction of FRBs are genuinely one-off is not ruled out by the data. One intriguing possibility is that a small fraction of FRBs might be associated with gravitational wave (GWs)
\citep{Totani:2013lia, Zhang:2013lta, Wang:2016dgs, Levin:2018mzg, Zhang:2020eou, Zhong:2020etx, Zhong:2019qoi}.
If confirmed, an association between an FRB and a GW event would reveal interesting physics before, during, and after compact binary mergers. 

Recently, a $2.8\sigma$ association between GW190425 (a high-mass binary neutron star merger) and FRB 20190425A was reported \citep{Moroianu:2022ccs}. The FRB was detected in the error region of the GW event 2.5 hours after the merger event. The dispersion measure of the FRB, which is an indicator of the source distance, is broadly consistent with the distance inferred from the GW data. A host galaxy was later identified for the FRB 
\citep{Panther:2022jfg},
whose redshift is also consistent with the GW event. 

To the first order, this association is fully consistent with the suggestion that the binary neutron star (BNS) merger left behind a supermassive neutron star, which collapses about 2.5 hours after the merger, as predicted by \citep{Zhang:2013lta} based on the so-called blitzar mechanism to produce FRBs \citep{Falcke:2013xpa}. Moroianu et al. \cite{Moroianu:2022ccs} interpreted the association and discussed the implications for the neutron star equation of state based on this scenario.
However, a subsequent scrutiny of the scenario led to some inconsistencies. First, the redshift of the host galaxy requires the GW event to be viewed at a large viewing angle much greater than $30^\circ$ in order for the FRB to escape the surrounding ejecta \cite{Bhardwaj:2023avo}. Second, deep searches left a tight upper limit on the brightness of the underlying kilonova, which essentially rules out the existence of an underlying supermassive magnetar with an initial spin period of $\sim 1 \, \rm{ms}$
\citep{Radice:2023zfv}.
Consequently, the GW190425/FRB 20190425A association is now placed on hold and generally regarded as a chance coincidence event by the community. 

In this paper, we propose an alternative mechanism to make a physical connection between GW190425 and FRB 20190425A within the framework of the Gertsenshtein–Zel'dovich (GZ) mechanism \citep{Gertsenshtein1962, Zeldovich1974}, that GWs propagating through a strong magnetic field induce electromagnetic (EM) radiation. {In our case, the kHz GWs are converted into kHz EM radiation via the GZ effect, which is then boosted to the GHz FRB emission via inverse Compton scattering (ICS) by relativistic particles.}

\begin{figure*}[ht]
    \centering
    \includegraphics[keepaspectratio, width=12cm]{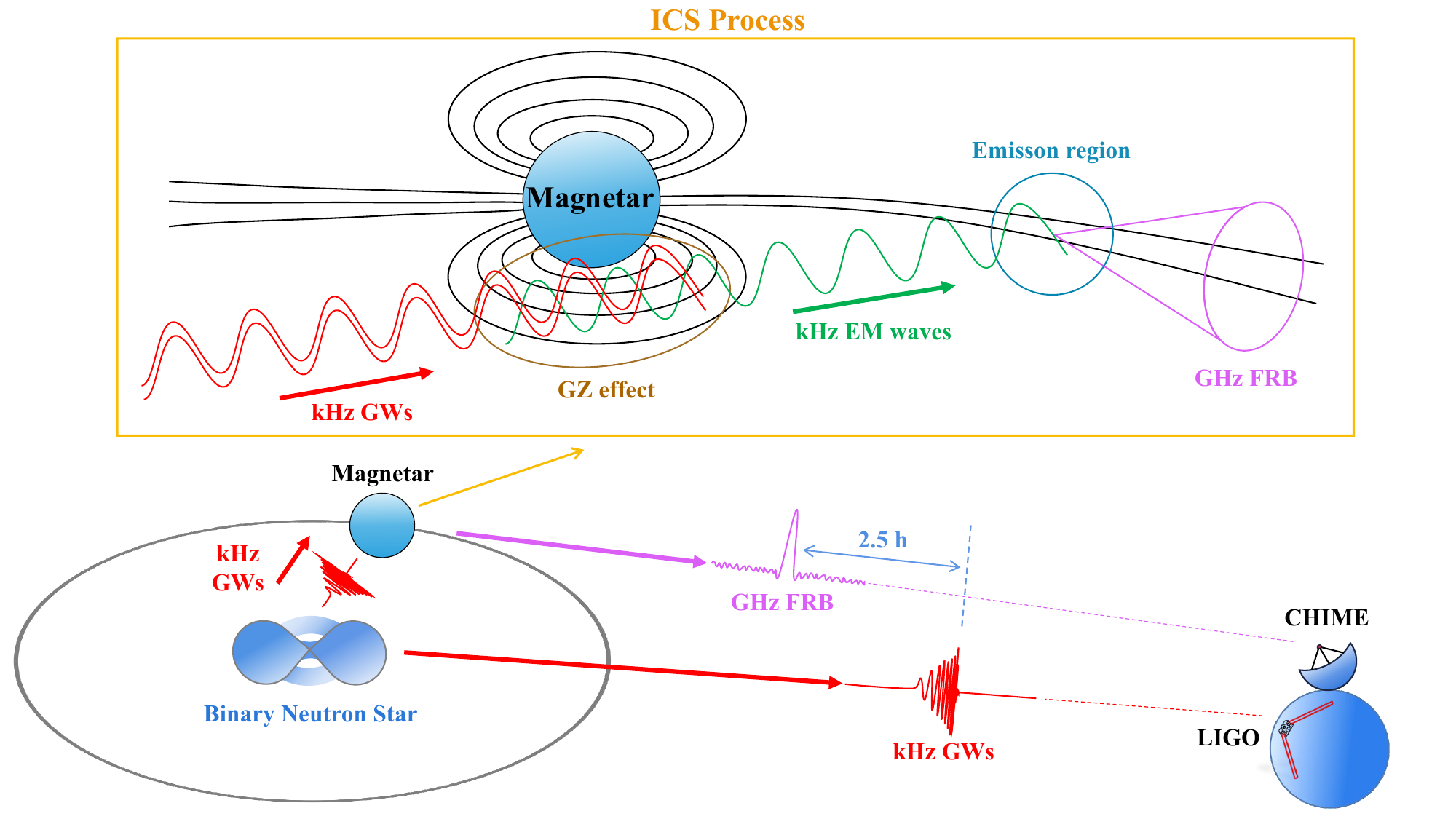}
\caption{The schematic diagram depicts the GW-FRB  association. In the lower panel, the BNS merger produces kHz GWs, with one component propagating directly to Earth for detection. The other component propagates a distance of $\sim 18 \, \rm{AU}$ to a magnetar. The upper panel details the magnetospheric physics: kHz GWs traverse the magnetar's magnetosphere, undergo GZ effect conversion to kHz EM waves, and then the frequency of EM waves is boosted to GHz via the ICS process to generate the FRB. This explains why FRB 20190425A arrives 2.5 hours after GW190425.
}
    \label{ICS}
\end{figure*}

\section{Conversion of GWs to EM waves}
In the presence of a strong magnetic field, GWs and EM waves can be converted to each other through the GZ effect. Recent studies suggest that a fraction of FRBs may originate from high frequency GWs via this mechanism when the GWs traverse the magnetosphere of a highly magnetized neutron star~\cite{Kushwaha:2022twx, Kalita:2022uyu}. In this work, we focus on the kHz frequency range, which is characteristic of GWs emitted during BNS mergers. 

The GWs strain scales intensity with the distance as $h \sim 1/D$. The distance, corresponding to the 2.5-hour delay time between the GW190425 and FRB 20190425A, is about $18 \, \rm{AU}$.
Given this, we adopt 
$|h_0| \sim 2.1 \times 10^{-10}\left(\frac{D}{18 \, \text{AU}}\right)^{-1}$ as the input value of the strain for our calculations, which is based on the relative configuration of the triple star system. 
The energy density of the GWs is given by \citep{misner1973gravitation,Maggiore:2007ulw}
\begin{align}
\rho_{\rm{GW} }  \simeq &\frac{c^2 \omega_0^2}{16 \pi G } \,\,  |h_0|^2 \approx 
3.0 \times 10^{15} \, {\rm erg/cm^3} \nonumber\\&\times \left(\frac{f_0}{2.5 \, \rm{kHz}}\right)^{2} \left(\frac{D}{18 \, \text{AU}}\right)^{-2}
\ , \end{align}
where $c$ denotes the speed of light,
 $G$ denotes the gravitational constant,
and $f_0 =\omega_0 / 2 \pi $ is the frequency for the GWs.

Following the derivation in \cite{Kushwaha:2022twx},
we obtain the energy density of the converted EM waves (see Appendix for the details):
\begin{align}
\label{eq:energy density of electromagnetic waves-magnetar}
\rho_{\rm{EM}} = & \alpha \rho_{\rm{GW} }  \simeq  \frac{B_0^2 |h_0|^2}{64 \pi } \frac{\reff}{\rthree^6} \approx  1.3 \times 10^{8} \, {\rm erg/cm^3}  \nonumber \\& \times
 \left(\frac{D}{18 \, \text{AU}}\right)^{-2} 
\left(\frac{B_0}{6.1\times 10^{15} \, {\rm G}}\right)^2 
\left(\frac{{\rthree}}{2}\right)^{-6}
\eta  \ .
\end{align}
Here, $\alpha  = \frac{G B_0^2}{4c^2 \omega_0^2 \rthree^6}  \reff $ is the conversion factor \cite{Kushwaha:2022twx}, $B_0$ is the surface magnetic field strength of the magnetar, ${\rthree} \equiv r/R_0$ is a dimensionless parameter for the conversion region, and $R_0 \sim 10 \, {\rm km}$ is the magnetar radius. 
$\reff$ is a dimensionless parameter defined as $\reff \equiv \left(1 + \xi^2 \omega_B^2 t^2  \right)$, $\omega_B$ is the rotation frequency of the magnetar, $t$ is the travel time of the GWs through the magnetosphere. The magnetic field perturbation ratio is defined as $\xi \equiv |\delta B| / {|B|}$, and is taken to be $0.01$ \cite{Pons:2012cm}. Since $\xi^2 \omega_B^2 t^2 \ll 1$, we ignore the correction and take $\eta\simeq 1$ in the following calculation. The magnetospheric EM oscillations in a magnetized plasma propagate via two distinct modes: the X-mode and the O-mode \cite{Zhang:2022uzl}. 
Whereas the propagation of O-mode is subject to many constraints, the X-mode can propagate freely similar to electromagnetic radiation. 

The energy budget of electromagnetic waves induced by gravitational waves is not large enough to power an FRB. Taking the strain of GW as $|h_0| \sim 2.1 \times 10^{-10}$. The energy that the magnetar absorbs from the GWs is $E_{\rm abs}= \epsilon J_{\rm GW} \sigma_{0} T_{\rm abs}\sim 10^{37}\,{\rm erg}$, where $J_{\rm GW}=c\rho_{\rm GW}$ is the energy flux of the GWs, $\sigma_{0}=4\pi R_0^2$ is the geometrical cross section of the magnetar. $T_{\rm abs}\sim 0.1\,{\rm s}$ is the absorption time, and $\epsilon \sim 10^{-2}$ represents the energy absorption ratio. This is much smaller than the typical energy~\cite{Wang:2017agh} released by the starquake $\delta E_{\rm cru}\sim 10^{44}\,{\rm erg}$. On the other hand, the GWs generated during the BNS merger would also trigger a f-mode resonant oscillation of the neutron star crust, which may eventually trigger a starquake at the magnetar surface \cite{Ho:1998hq,Yu:2024uxt}. Such an explosion may excite a relativistic outflow in the magnetosphere and accelerate relativistic particles, which would upscatter the low-frequency X-mode electromagnetic waves to power an FRB \cite{Zhang:2021pfn,Qu:2024cjh}.

\section{Inverse Compton Scattering}

The GWs produced by BNS mergers typically exhibit frequencies in the range of hundreds to thousands of Hertz. The converted EM waves also lie within this frequency band. The characteristic frequency of FRBs spans from 100 MHz to 8 GHz. 
One mechanism to resolve such a frequency discrepancy is through the inverse Compton scattering (ICS) mechanism proposed in \cite{Zhang:2021pfn,Qu:2024cjh}. In the following, we reiterate some of the key points of such a model. 

An ICS process can boost kHz frequency EM signals to the GHz band \cite{Zhang:2021pfn,Qu:2024cjh}:
\begin{align}
f_{\rm FRB} \simeq \gamma^2 f_0 
\approx 625 \, {\rm MHz} \left(\frac{\gamma}{500} \right)^2 \left(\frac{f_0}{2.5 \, {\rm kHz}} \right)\ .
\end{align}
Here, $\gamma$ is the Lorentz factor of relativistic particles, and $f_0$ is the frequency of the incident EM waves induced by GWs, which we normalize to $ f_0 =2.5 \, {\rm kHz}$. 
 
The ICS emission power of a single relativistic lepton (electron or positron) can be calculated as $P_{\rm ICS}  = \gamma^2 \sigma_{\rm X}c \rho_{\rm EM}  \approx  5 \times 10^{-17} \,  {\rm erg / s} \left(\frac{{\rone}}{500}\right)^6  \left(\frac{{\rthree}}{2}\right)^{-6} \left(\frac{D}{18 \, \text{AU}}\right)^{-2}$, where $\sigma_{\rm X}  \sim \omega_{\rm{c}}^{-2} \sim B_0^{-2}\rone^{6} $ is the cross section of the Compton scattering in the presence of a strong magnetic field, $\omega_{\rm{c}}$ is the cyclotron frequency of the charged particle,  ${\rone} \equiv r/R_0$ is a dimensionless parameter for the scattering region. The derivation of $\sigma_{\rm X}$ is detailed in \cite{Qu:2024cjh}. 
Within the magnetar's magnetosphere, low frequency EM waves may be able to redistribute relativistic leptons and modulate the particle density to form bunches, with the spatial scales of these bunches matching the wavelengths observed in FRBs. A preliminary scenario for such a bunching mechanism has been discussed in \cite{Qu:2024cjh}. Within the ICS picture, the magnetar's outer magnetosphere is expected to be charge-depleted, with a parallel electric field balancing rapid cooling of the bunches.  See detailed discussion in \cite{Qu:2024cjh}.

{In the emission region, the coherent radiation power depends on the net charges per bunch $N_{e,b}= \frac{\pi \zeta B_0 }{q c T} \left(\frac{f_{\rm FRB}}{c}\right)^{-3} \rtwo^{-3} \gamma^{2}$ and the total number of bunches $N_{b}= 
\left(\frac{f_{\rm FRB}}{c}\right)^{2}
{(R_0 \rtwo)^{2}}\gamma^{-3}$ \cite{Qu:2024cjh}.}
Here, ${\rtwo}$ is a dimensionless parameter for the emission region, defined as ${\rtwo} \equiv r/R_0$, $q$ is the charge of the charged particle (electron or positron), $\zeta$ is the net charge factor, and $T$ is the magnetar period. 
The total luminosity via the ICS process is given by $L_{\rm {ICS}} \simeq \gamma^4 N_{b}^2N_{e,b}^2 P_{\rm {ICS}} =\gamma^6 N_{b}^2N_{e,b}^2  \sigma_{\rm X} c \rho_{\rm EM}
$, and the final result is 
\begin{align}
L_{\rm {ICS}} \approx  &
2.2 \times10^{41} \, {\rm {erg/ s}}  \left(\frac{B_0}{6.1 \times 10^{15} \, \rm{G}}\right)^2 \left(\frac{D}{18 \, \text{AU}}\right)^{-2} \nonumber \\
     & \times  
    \left(\frac{{\rthree}}{2}\right)^{-6}\left(\frac{{\rone}}{500}\right)^6   \left(\frac{{\rtwo}}{500}\right)^{-2}
    \left(\frac{T}{1\text{s}}\right)^{-2}
    \zeta^2 .
  \label{eqTOT}
\end{align}
The dimensionless radial parameters $\rthree$, $\rone$ and $\rtwo$ correspond to the GZ conversion, the scattering and the emission regions (in terms of the magnetar radius $R_0$), respectively.
We can obtain the flux density $ S  =  \frac{L_{\rm ICS}}{4\pi d^2 \Delta \nu} 
\approx 19 \, {\rm Jy} \left(\frac{\Delta \nu}{400 \, {\rm MHz}}\right)^{-1} \left(\frac{L_{\rm ICS}}{2.2 \times10^{41} \, {\rm erg/ s}} \right) \left(\frac{d}{156 \,  {\rm Mpc}}\right)^{-2} $.
Here, $d$ is defined as the distance from the source to the Earth and $\Delta \nu$ is the bandwidth of the FRB~\cite{Zhang:2022uzl}.

Thus, we establish a possible physical connection between GW190425 and FRB 20190425A: In a system where a magnetar exists in the vicinity of a BNS merger, possible in a hierarchical triple system, the BNS merger generates GWs, which propagate $18 \, \rm AU$ ($\sim$2.5 hours) to encounter the magnetar. Subsequently, part of GWs are converted to EM radiation via the GZ effect. The GWs also trigger crustal quakes to launch an explosion similar to regular magnetars. The low-frequency EM radiation is boosted to FRB frequency through the ICS mechanism, resulting in FRB 20190425A. Since the FRB emission mechanism is similar to those of traditional magnetar-generated FRBs, we do not expect the observational properties of FRB 20190425A to be much different from other FRBs. This is consistent with the observations of FRB 20190425A. 

\vspace{-5pt}
\section{Conclusions and discussion}
In this work, we propose a novel mechanism to explain the association between GW190425 and FRB 20190425A. We suggest that GWs emitted during the BNS merger are converted into kHz EM waves via the GZ effect near the magnetar. The same GWs also trigger neutron star crustal quakes which launch energetic relativistic outflows. The kHz EM waves undergo ICS with the relativistic charged bunches and get boosted to high frequencies and produce the observed FRB signal. This model provides a plausible physical link connecting GW190425 and FRB 20190425A and offers new insight into the origin of FRBs. 

Within this scenario, a rare fraction of FRBs may be associated with GWs not only for BNS mergers, but also binary black hole mergers, with the FRB signal delaying from the GW signal by a timescale determined by the distance between the GW source and the putative magnetar. The event rate density of these FRBs is very difficult to estimate, depending on the likelihood of the existence of a magnetar in the vicinity of the GW sources. Since the GW event rate density is much smaller than that of FRBs, we expect that FRBs produced by this mechanism are much rarer than normal FRBs. Since the radiation mechanism of FRBs of this type is similar to traditional FRBs, these GW-associated FRBs may not be observationally distinguishable from most FRBs. Also, it is expected that most GW sources are not followed by FRBs. In any case, systematically searching for GW-FRBs associations with FRBs lag GWs by hours will be meaningful and will play interesting constraints on such a scenario. Since there are more BNS mergers than BBH mergers in the universe, the probability of detecting such associations with BNS mergers should be higher than those with BBH mergers, which is consistent with the putative GW190425/FRB 20190425A association.

\appendix


\section{Appendix A: The Gertsenshtein-Zel'dovich effect}
\label{append_GZ}

According to the GZ effect \citep{Gertsenshtein1962,Zeldovich1974}, gravitational waves (GWs) propagating through a background magnetic field induce electromagnetic (EM) fields, leading to the partial conversion of GWs into EM radiation.
The action for the EM fields in curved spacetime can be expressed as
\begin{align}
    S=\int d^4x \sqrt{-g} \left(\frac{1}{16\pi G}R - \frac{1}{4} F^{\mu\nu} F_{\mu\nu}
    \right).
\end{align}
Focusing on the induced EM waves, we apply the variational principle to the 4-potential $A_\mu$ and derive
$\delta S= \int d^4x \sqrt{-g}F^{\mu\nu}\partial_\nu \delta A_\mu
=-\int d^4x \partial_\nu\left(\sqrt{-g}F^{\mu\nu} 
    \right)    \delta A_\mu $,
where integration by parts is employed. Therefore, the equations of motion for the EM fields are expressed as
\begin{align}\label{eomEM}
    \partial_\nu
    \left(\sqrt{-g}F^{\mu\nu} 
    \right)=0\ .
\end{align}

For the curved spacetime generated by GWs, the metric takes the form $g_{\mu\nu}=\eta_{\mu\nu}+h_{\mu\nu}$, where $\eta_{\mu\nu}$ is the Minkowski metric, and $h_{\mu\nu}$ represents the perturbation induced by GWs. The determinant of the metric is $g=-1-\eta^{\mu\nu}h_{\mu\nu}+\mathcal{O}(h^2)$, so we obtain
\begin{align}
    &{\partial _\nu }\left(\sqrt { - g} {F^{\mu \nu }}\right)
    \simeq \ {\partial _\nu }
    \left({F^{\mu \nu }} + \frac{1}{2}h{F^{\mu \nu }}\right)\nonumber\\
    \simeq\  &{\partial _\nu }\left({F^{\mu \nu }_{(0)}} + \frac{1}{2}h{F^{\mu \nu }_{(0)}} + {h^\nu }_\rho {F^{\rho \mu }_{(0)}} - {h^\mu }_\rho {F^{\rho \nu }_{(0)}} \right)\ ,
\end{align}
where $h=\eta^{\mu\nu}h_{\mu\nu}$ and $F_{(0)}^{\mu \nu } = {\eta ^{\mu \rho }}{\eta ^{\nu \sigma }}{F_{\rho \sigma }}$ (see, e.g.~\cite{Plebanski:1959ff,Aggarwal:2020olq,Palessandro:2023tee,Hwang:2023nqx}). Thus, the Maxwell's equations in Eq.~\eqref{eomEM} become
\begin{align}
    \partial^\nu\left(F_{\mu\nu}+\frac{1}{2}h F_{\mu\nu}+h^\rho_{\ \nu}F_{\rho\mu}-h^\rho_{\ \mu}F_{\rho\nu}
    \right)=0\ .
    \label{eq_eom1}
\end{align}
We take the propagation direction of the GWs to be along the $z$-axis. In the transverse-traceless (TT) gauge, the $i,j$ component of the metric perturbation takes the form
\begin{align}
    h^{TT}_{ij}=&\ h_+e^+_{ij}
    +h_\times e^\times_{ij}\ ,
    \label{hTT}
\end{align}
where
\begin{align}
    h_+ &= |h_+| e^{i\left(k_{0} z - \omega_0 t\right)},\  h_\times = |h_\times| e^{i\left(k_{0} z - \omega_0 t+\Theta\right)},
\end{align}
and $\Theta$ denotes the phase difference between the two polarizations. The polarization tensors $e^+_{ij}$ and $e^\times_{ij}$ are
\begin{align}
e^+_{ij} 
&= \begin{pmatrix}
 1 & 0 & 0 \\
 0 & -1 & 0\\
 0 & 0 & 0  \end{pmatrix},\ \ e^\times_{ij} 
=\begin{pmatrix}
 0 & 1 & 0 \\
 1 & 0 & 0\\
 0 & 0 & 0  \end{pmatrix}.
\end{align}

\subsection{Perturbation case}
 
We consider a time-varying magnetic field oriented along the $y$-axis: $\mathbf{B}(t) = \left(0, |B| + |\delta B|\sin(\omega_B t), 0\right)$ \citep{Kushwaha:2022twx, Kalita:2022uyu}, where $\omega_B$ is the pulsar rotation frequency.  Due to Faraday's law $\tfrac{1}{c}\tfrac{\partial\mathbf{B}}{\partial t} +\nabla\times\mathbf{E}=0$, this magnetic field would induce an electric field $ \mathcal{E}_x = - (z \, |\delta B| \omega_B/{c}) \cos(\omega_B t)
$. After taking into account the EM waves induced by the GWs, the EM field tensor is given by \citep{Kushwaha:2022twx, Kalita:2022uyu}
\begin{align}
  \label{eq:EMField-tensor-dd_matrix}
  F_{\mu\nu} = \mathcal{F}_{\mu\nu} + \hat{F}_{\mu\nu}\ .
\end{align}
Here the background field tensor is 
\begin{align}
  \label{eq2}
  \mathcal{F}_{\mu\nu} = \begin{pmatrix}
  0 & \mathcal{E}_x & 0 & 0\\
  -\mathcal{E}_x & 0 & 0 & \mathcal{B}_y \\
  0 & 0& 0 & 0\\
  0 & - \mathcal{B}_y & 0 & 0  
  \end{pmatrix}\ ,
\end{align}
where $\mathcal{B}_y = |B|+ |\delta B| \sin (\omega_B t)$, and $\hat{F}_{\mu\nu}$ is the field tensor induced by the GWs
\begin{align}\label{eq3}
\hat{F}_{\mu\nu} 
= \begin{pmatrix}
0 & \hat{E}_x & \hat{E}_y & \hat{E}_z \\
-\hat{E}_x & 0 & -\hat{B}_z & \hat{B}_y \\
-\hat{E}_y & \hat{B}_z & 0 & -\hat{B}_x\\
-\hat{E}_z & - \hat{B}_y & \hat{B}_x & 0  \end{pmatrix}.
\end{align}
To determine the EM waves induced by the GWs, we plug Eq.~\eqref{hTT} into  Eq.~\eqref{eq_eom1}.
To first order in $h_{\mu\nu}$ and using Faraday's law, we can obtain
\begin{align}
    &\frac{1}{c} \partial_t \hat{E}_x - \partial_y \hat{B}_z +\partial_z \hat{B}_y = - \frac{1}{c} \partial_t \left(\mathcal{E}_x h_+ \right) -\partial_z (\mathcal{B}_y  h_+) 
    \ ,
    \nonumber \\&
  \frac{1}{c} \partial_t\hat{B}_y - \partial_x\hat{E}_z + \partial_z\hat{E}_x = 0 \ .
\end{align}
Given that both GWs and EM waves propagate along the $z$-axis, we set $\hat{E}_z = \hat{B}_z = 0$. Thus, we can obtain
\begin{subequations}
\begin{align}\label{eq:WaveEquation-Ex-FF}
\frac{1}{c^2} \frac{\partial^2  \hat{E}_x }{\partial t^2} - \partial_z^2 \hat{E}_x  
&= - f_{E}(z,t)\ , \\
\label{eq:WaveEquation-By-FF}
\frac{1}{c^2} \frac{\partial^2  \hat{B}_y }{\partial t^2} - \partial_z^2 \hat{B}_y 
&= - f_{B}(z,t)\ ,
\end{align}
\end{subequations}
where $f_{E}(z,t)$ and $f_{B}(z,t)$ are the sources induced by GWs:
\begin{subequations}
   \begin{align}
&f_{E}(z,t) =|h_+| {|B|}\frac{k_0 \omega_0}{c}  
e^{i (k_0 z - \omega_0 t) }\   \nonumber \\
& + \frac{z |h_+| |B| \xi  \omega_B }{2 c^3}  \left[ \omega_+^2 e^{- i \omega_+ t}  +  \omega_-^2  e^{-i \omega_- t} \right]e^{i k_0 z}   \nonumber\\
&+
\frac{i |h_+| |B| \xi  k_0 }{2 c}  
\left[ \omega_+ e^{- i \omega_+ t}  -  \omega_-  e^{-i \omega_- t} \right]e^{i k_0 z},
 \\
& f_{B}(z,t) = 
|h_+| {|B|}k_0^2 e^{i \left(k_0z - \omega_0 t \right)}
 \nonumber \\
& + \frac{ |h_+| |B| \xi  \omega_B }{2 c^2}  (z k_0-i)\left[ \omega_+ e^{- i \omega_+ t}  +  \omega_-  e^{-i \omega_- t} \right]e^{i k_0 z}\nonumber\\
&+\frac{i |h_+| |B| \xi  k_0^2 }{2} \left[ e^{-i \omega_+ t  }  - e^{-i  \omega_- t } \right]e^{i k_0 z}.
\end{align}
\end{subequations}
Here, $\xi \equiv |\delta B| / {|B|}$, and
$\omega_{\pm} = \omega_0 \pm \omega_B$. We assume that $|h_+|=|h_\times|=h_0$. The solutions of Eqs. (\ref{eq:WaveEquation-Ex-FF}) and (\ref{eq:WaveEquation-By-FF}) can be solved as
\begin{align}
    \hat{E}_x & \simeq - \frac{h_0}{4} {|B|} \left(1 - \xi  \omega_{B} t\right)e^{i\left(k_0 z - \omega_0 t\right)}\ ,\label{EM1} \\
    \hat{B}_y & \simeq \frac{h_0}{4}{|B|} \left(1 + \xi \omega_B t\right) e^{i \left(k_0 z - \omega_0 t\right)}\ .\label{EM2} 
\end{align}
Note that this result is obtained under the condition: $\omega_0 \sim \text{kHz} \gg \omega_B$. In magnetars, the characteristic frequency $\omega_{B}$ of the time-varying magnetic field is approximately 1 Hz \citep{Kushwaha:2022twx}. 

The energy density of GWs can be expressed as \citep{misner1973gravitation}
\begin{align}\label{eq:rho_GW}
\rho_{\rm{GW} } = \frac{c^2 \omega_0^2}{32 \pi G } \left(  |h_+|^2 + |h_\times|^2 \right) = \frac{c^2 \omega_0^2}{16 \pi G } \,\,  |h_0|^2 \ .
\end{align}
For the strength of the magnetic field, it follows the distribution $|B|=B_0(r/R_0)^{-3}=B_0/\rthree^3$ near the magnetar~\cite{Goldreich:1969sb}, where $B_0$ is the magnetic field strength at the surface of the magnetar, $R_0 \sim 10 \, \rm{km}$ is the magnetar radius and $\rthree$ is the conversion region. With Eqs.~\eqref{EM1} and \eqref{EM2} , the energy density of the induced EM waves \cite{jackson1998classical}  is calculated to be
\begin{align}
    \rho_{\rm{EM}} & = \frac{|{\hat{E}_x}|^2 + |\hat{B}_y|^2 }{8 \pi} 
\simeq \frac{ B_0^2 |h_0 |^2}{64 \pi} \frac{\reff}{\rthree^{6}}   \  ,
\end{align}
where the dimensionless parameter:
\begin{align} 
\reff &   =1 + \xi^2 \omega_B^2 t^2 \ .
\end{align}
Here, $t$ is the travel time of the GWs through the magnetosphere. Given that $\xi^2 \omega_B^2 t^2 \ll 1$, it can be neglected as a small quantity in calculations and we take $\eta \simeq 1$.
The conversion efficiency can be written as
\begin{align}
    \label{eq:alpha-point}
    \alpha \equiv  & \frac{ \rho_{\rm EM } }{\rho_{\rm GW }} = \frac{G B_0^2}{4c^2 \omega_0^2 \rthree^{6}}  \reff
    \approx \left(4.4 \times 10^{-8}\right) 
    \nonumber \\&  \times
    \left(\frac{B_0}{6.1 \times 10^{15} \, {\rm G}} \right)^{2}
  \left(\frac{f_0}{2.5 \, \rm{kHz}}\right)^{-2}   \left(\frac{{\rthree}}{2}\right)^{-6}
    \eta \ .
\end{align}
Here, $f_0 =\omega_0 / 2 \pi $ is the frequency for the GWs.

\section{Appendix B: The scattering cross section of ICS}
\label{Appendix ICS}

To derive the cross section for X-mode waves in a magnetized plasma, we employ the classical framework, beginning with the dynamics of a charged particle (electron or positron) in EM fields \cite{Qu:2024cjh}, while a quantum-mechanical treatment has been discussed in \cite{Herold:1979zz}. It is noteworthy that all computations presented in this appendix are performed in the comoving frame.
In the background fields $\mathbf{B'}(t') = \left(0, |B'| + |\delta B'|\sin(\omega_B' t), 0\right)$ 
and $\mathbf{E'}(t') = (E'_0 e^{-i \omega_0'  t'},0,0)$, a particle with mass $m_e$, charge $q$, and velocity $\mathbf{v'}$ obeys
\begin{align}
    \label{charged particle following Nt.eq}
    m_e \frac{d\mathbf{v'}}{dt'} = q(\mathbf{E'}+\mathbf{v'} \times \mathbf{B'})\ .
\end{align}
The velocity of the charged particle can be solved as
\begin{align}
    \label{The velocity of charge particle}
    \mathbf{v'} \simeq \frac{i q \omega_0' E'_0 e^{-i \omega_0' t'}}{m_e (\omega_0'^2 - \omega_{\rm{c}}'^2)}\vec{x} - \frac{q \omega_{\rm{c}}' E'_0 e^{-i \omega_0' t'}}{m_e (\omega_0'^2-\omega_{\rm{c}}'^2)}\vec{z}\ ,
\end{align}
where $\omega_{\rm{c}}'$ is the cyclotron frequency of the charged particle. It is worth noting that Eq.(\ref{The velocity of charge particle}) presents an analytical solution obtained by neglecting the $|\delta B'|$, which is a small parameter. In addition, $\omega_{\rm{c}}$ is the cyclotron frequency in the lab frame and $\omega_{\rm{c}} = \omega_{\rm{c}}'$   \cite{Qu:2024cjh}, we can obtain
\begin{align}
   \omega_{\rm{c}} = \frac{q|B|}{m_ec}   \approx & (8.6 \times 10^{14} \, {\rm rad / s}) \left(\frac{\rone}{500}\right)^{-3} 
   \left(\frac{B_0}{6.1 \times 10^{15} \, {\rm G}}\right),
 \end{align}
where $q = e$ and $m_e = 0.511 \, \rm{MeV/c^2}$.
Near the magnetar, $|B|=B_0(r/R_0)^{-3}=B_0/\rone^3$ and $\rone$ represents the scattering region. 

The radiation is induced by the motion of the charged particle \cite{jackson1998classical}, which can be written as $\mathbf{E'_{\rm {rad}}} = \frac{q}{c^3 r'} \left[ \mathbf{n'} \times \left( \mathbf{n'} \times  \dot{\mathbf{v}}' \right)\right]$ and $\mathbf{B'_{\rm {rad}}}  = \frac{1}{c}  \mathbf{n'} \times \mathbf{E'_{\rm {rad}}}$.
Then we can calculate the cross section: $\sigma' = \frac{P'}{S'} = (E'^2_{\rm {rad}} r'^2) / {E'^2_0}$,
where $P'$ is the emitted power and $S'$ is the incident Poynting flux in the comoving frame. 
Using these equations, we can obtain the scattering cross section of X-mode in the presence of a strong magnetic field 
\begin{align}
    \label{cross section x-mode in lab frame}
    \sigma'_{\rm {X}} = \frac{\sigma_{\rm T}}{2}\left [ \frac{{\omega_0'}^2}{(\omega_0'+ \omega_{\rm{c}})^2} +  \frac{{\omega_0'}^2}{(\omega_0'- \omega_{\rm{c}})^2} \right ]\ ,
\end{align}
where $\sigma_{\rm T}=\tfrac{8\pi}{3}\left(\tfrac{q^2}{m_ec^2}\right)^2$ is the Compton scattering cross section. In order to connect the lab frame and the comoving frame, it is necessary to use the Lorentz transformations $\sigma_{\rm X} = (1-\beta\cos\theta_i) \sigma'_{\rm X}$. Based on this, we obtain the scattering cross section of the X-mode in the lab frame: 
\begin{align}
    \label{cross section x-mode in lab frame2}
    \sigma_{\rm X} & = \nonumber \frac{\sigma_{\rm T} }{2} \left(1-\beta\cos\theta_i \right)  \left [ \frac{\omega_0'^2}{(\omega_0'+\omega_{\rm{c}})^2} +  \frac{\omega_0'^2}{(\omega_0'-\omega_{\rm{c}})^2} \right ] \\
         &\approx (5.3\times 10^{-41} \, {\rm cm}^2)  \left(\frac{\rone}{500}\right)^{6} 
   \left(\frac{B_0}{6.1 \times 10^{15} \, {\rm G}}\right)^{-2} \ .
\end{align}
It should be noted that $\omega_0' = \gamma \left(1-\beta\cos\theta_i \right)\omega_0 \sim \gamma \omega_0$ for $\theta_i > 0$. Additionally, in the highly magnetized magnetosphere, we have $\omega_{\rm{c}} \gg \omega_0' \sim \rm{MHz}$.

\section*{Acknowledgments}
This work is supported by the National Natural Science Foundation of China (No.12588101, No.12375059).
We are grateful to Qing-jie Yuan and Zanothando Kubheka for discussions related to the GZ effect, and to Yuanhong Qu for discussions on the ICS mechanism.

\vspace{-10pt}

\bibliography{2603v3}
\footnotesize

\end{document}